# Investigating Exit Choice in Built Environment Evacuation combining Immersive Virtual Reality and Discrete Choice Modelling

Ruggiero Lovreglio[a], Elise Dillies[b], Erica Kuligowski[c], Anass Rahouti[b,d], Milad Haghani[e]

[a] School of Built Environment, Massey University, Auckland, New Zealand
[b] Department of Civil Engineering and Structural Mechanics, University of Mons, Mons, Belgium
[c] School of Engineering, RMIT University, Melbourne, Australia
[d] Fire safety consulting sprl, Ligny, Belgium
[e] School of Civil and Environmental Engineering, The University of New South Wales, UNSW Sydney, Australia

**Abstract:**
In the event of a fire emergency in the built environment, occupants face a range of evacuation decisions, including the choice of exits. An important question from the standpoint of evacuation safety is how evacuees make these choices and what factors affect their choices. Understanding how humans weigh these (often) competing factors is essential knowledge for evacuation planning and safe design. Here, we use immersive Virtual Reality (VR) experiments to investigate, in controlled settings, how these trade-offs are made using empirical data and econometric choice models. In each VR scenario, participants are confronted with trade-offs between choosing exits that are familiar to them, exits that are less occupied, exits that are nearer to them and exits to which visibility is less affected by fire smoke. The marginal role of these competing factors on their decisions is quantified in a discrete choice model. Post-experiment questionnaires also determine factors such as their perceived realism and emotion evoked by the VR evacuation experience. Results indicate that none of the investigated factors dominated the others in terms of their influence on exit choices. The participants exhibited patterns of multi-attribute conjoint decision-making, consistent with the recent findings in the literature. While lack of familiarity and the presence of smoke both negatively affected the desirability of an exit to evacuees, neither solely determined exit choice. It was also observed that prioritisation of the said factors by participants changed during the repeated scenarios when compared to the first scenario that they experienced. Results have implications for both fire safety designs and future VR evacuation experiment designs. These empirical models can also be employed as input in computer simulations of building evacuation.

**Keywords:** exit choice; fire evacuation; discrete choice; random utility; virtual reality

## 1. Introduction

Understanding human behaviour in building fires is an important part of fire safety design [1]. Engineers are tasked with ensuring that a building design provides a sufficient level of safety for its occupants under a variety of different fire and evacuation scenarios [2]. It is during the development and assessment of these scenarios that occupant behaviour must be appropriately considered, requiring theory and data on evacuation likelihood, pre-travel delay times or actions, route choice, movement attributes, such as walking speeds, and exit choice [3,4].

A cohort of data that exist on occupant behaviour during building fires has been collected from *Revealed Preference* (RP) studies [5–8]. These studies collect data on people's actual choices made, for example, during evacuation drills. While post-fire studies offer insights on real-world decisions and behaviours of evacuees, building evacuation drills provide data primarily focused on movement attributes and overall congestion points throughout the building. In both cases, researchers have limited experimental control over the factors under investigation, including the types of evacuees involved or the type of scenarios that they face (e.g., in terms of the architecture of the space or the level of crowding that they face, etc.) [9].

Evacuee data can also be collected by *Stated Preference* (SP) experiments [10–18]. SP data are collected by presenting participants with a hypothetical scenario and asking what they would do (i.e., their behavioural intention) in that particular scenario [19]. These data are often captured via survey designs. While SP designs give the researcher a high level of control over the factors under investigation, they have lower *ecological validity* given that subjects are aware that they are part of an experiment [19,20].

While laboratory experiments can also offer a high level of control over factors of interest, even these environments may not be realistic enough [21]. The use of emerging technologies, like virtual or augmented reality, to study human behaviour during building fires has increased [22–26]. These technologies allow researchers to gain a better understanding and observe the decision-making of known subjects in real-time and even inquire about behaviours in post-experiment surveys. Both immersive and non-immersive simulated environments also allow researchers to safely introduce and test conditions, like smoke, that would be impossible to achieve in real-world settings. Research has demonstrated that results obtained in VR settings are comparable to data from the "real world"; although further validation is required [27,28].

The aim of this work is to provide new insights on the exit choice behaviour of occupants during fires. To achieve this goal, we designed a new immersive virtual reality (VR) exit choice experiment to investigate how certain factors affect exit choice during a fire evacuation. The factors investigated were the behaviour of other evacuees, the presence of smoke, the distance of an exit and the decision maker's familiarity with an exit. VR was used in this study to obtain real-time data on exit choice during a simulated fire emergency as well as feedback from participants once the experiment was over. These data can be used by engineers in their fire safety analyses as qualitative inputs into the development of evacuation scenarios or as quantitative inputs into evacuation models.

## 2. Exit Choice Literature

A number of studies have been performed on exit choice during building fires [21]. These studies have identified the multiple factors that can influence evacuee exit choice, including familiarity, social influence, affordances and design of the built environment, and harmful conditions within the environment itself (e.g., smoke). For each of these factors, previous research findings will be further explored.

In fire emergencies, people have a tendency to evacuate via familiar routes and exits [29]. Proulx [30] found that people were drawn to familiar exits, and even in circumstances where other exits were available and/or closer, people were unlikely to adopt a new route or exit previously unknown to them during an emergency. Several studies of actual fire events (e.g., [31,32]), as well as evacuation drills (e.g., [33]), provide evidence for evacuee tendencies to gravitate toward the familiar. Experiments performed in a mock furniture store setting found that 71% of occupants chose the door they used to enter the building for evacuation as well [33]. Also, studies using VR settings of a hotel [34] and a museum [18] demonstrated that familiarity influenced exit choice. In the first case, people were more likely to evacuate via the main entrance unless provided with additional information about the buildings (in this case via signage), and in the second case, familiarity was enhanced when others also used the same exit and reduced when more people left through the other exit.

Other occupants can also influence an evacuee's choice of exit, and this influence can vary with the number of people present in the scenario. In very large crowds, 75 to 150 people as an example, participants were not observed following "the crowd" [35]. However, observations of studies involving a much smaller number of participants provide evidence of this trend. Limited to around 50 participants on a virtual train platform, Lin et al. [36] found that uneven splits of the other evacuees influenced participants to follow the majority; and this finding held true across multiple cultures. Three studies with similar designs investigated exit choices with crowd sizes up to 20 people. In the first two studies [11,13], participants viewed videos of the scenarios via an online survey and in the other, immersive VR technology was used [37]. In both cases, participants were confronted with a hypothetical situation and asked to choose between two exits. Watching the online videos, participants favoured less crowded exits except when all virtual agents used the same exit. In the VR study, participants were likely to follow smaller crowds (~10); and in cases of larger crowds (~20), only if the doors were wider in size.

Experimental studies with and without VR tested the influence of small numbers of people on exit choice behaviour. Investigating wayfinding from an enclosed room into a corridor, Zhu et al. [38] tested the influence of 1-3 actors instructed to choose a route that opposed the signage and found that in one-actor conditions, evacuees followed the actor's behaviour. Also, studying social influence in tunnel fires, studies have found that participants were more likely to move to the emergency exit [39] and along the shorter path to the exit [40] when a virtual agent within the simulation did so.

Finally, information and conditions within the built environment can also influence exit choice. Nilsson et al. [41], adapting Gibson's Theory of Affordances [42] to building fire emergencies, note that the attractiveness of an exit is related to how well it affords egress. One of these factors is functional affordance, in that the exit allows for safe and effective movement to safety (e.g., is clear of smoke). While research provides evidence that people do walk through smoke under varying obscuration levels [43–45], a study of route choice behaviour in a virtual building showed that heavy smoke can reduce the use of evacuation shortcuts. It should be noted; however, that the presence of others evacuating via that shortcut increased its use rate [46].

While a notable amount of work has been done looking individually at how these factors influence exit or route choice, only a few studies included multiple factors (e.g., [37]). Additionally, no study was found that included familiarity, social influence and smoke conditions in the same experiment to investigate their combined influence on exit choice. Based on the literature and the proposed factors of study in this work, the following three hypotheses were developed.

**Hypotheses:**

- People are more likely to use an exit that is more familiar to them rather than one that is less familiar, even if that exit is located farther away from them.
- People are more likely to follow others, and in turn, use an exit that has a larger number of people using it compared with the other exits that may be less populated.
- People are more likely to use exits that are free from smoke and other harmful conditions.

## 3. Material and Methods

This section provides a description of the VR experiment designed in this study (Section 3.1), the experimental procedure (Section 3.2), the participants who took part to the experiment (Section 3.3) and the statistical analysis tool used to investigate the participants' choices (Section 3.4).

### 3.1 VR Experiment Design

During the VR experiment, a participant is placed within a virtual room where he/she has to choose an exit to use to leave the room during a simulated fire emergency. As shown in Figure 1, the room has 3 exits (A, B, and C) from which the participant can choose. The geometry illustrates where the decision-maker was located at the start of the experiment (i.e., starting point), while the red square identifies the location where, when reached, the first event in the experiment was triggered (i.e., the fire evacuation emergency). Once a participant exited via one of the three provided exits, the experiment ended.

Within the experiment, the available exits differed by a range of physical, social and individual factors. Depending upon the scenario, the exits differed by their physical distance away from the decision-maker and the presence of smoke (entering from the top of some of the exits). Additionally, the exits presented to the participants differed by the number of evacuees who were already using each exit (i.e., social factor) and how familiar the exit was to the participant (i.e., individual factor). Familiarity was defined as an exit that the participant had used in the past; and since one of the exits (Exit A) was located along the path participants needed to travel to reach their starting point in the experiment (see the red square in Figure 1), it was assumed to be an exit that was more familiar than the others.

In the next section of the paper, the following abbreviations are used to identify the variables under investigation:

- NP: Number of people using an exit;
- DIST: Distance of the participant from an exit;
- SMOKE: Presence of smoke;
- FAM: Familiarity of the participant with the exit.

Each of the four variables above has a number of dimensions that can be varied within the experiment. For simplicity purposes, the following assumption was made for the NP variable: the number of evacuees leaving the room by each exit (NP) can be equal to 0, 1, 5 or 10 (4 levels/dimensions of the variable per exit). Additionally, the distance of the decision-maker from the exits was defined using the geometry of the virtual environment. While the location of Exit A was kept constant (6.0 m), the location of Exits B and C varied across scenarios. See Figure 1 for the two positions possible for both Exits B and C.

The distance of the participant from Exit B was either 3.6 m or 5.6 m, while the distance to Exit C was 3.0 m or 4.6 m. For each exit, smoke (SMOKE) was either present, entering from the top of the door during the scenario or not. The value of this variable was set to 1 when smoke was present and 0 otherwise. The final constraint for the scenario development was that participants were only familiar with Exit A as they accessed the room only using this door (i.e., 1 familiar and 0 non-familiar) at the beginning of the experiment. The summary of the levels for each variable is shown in Table 1.

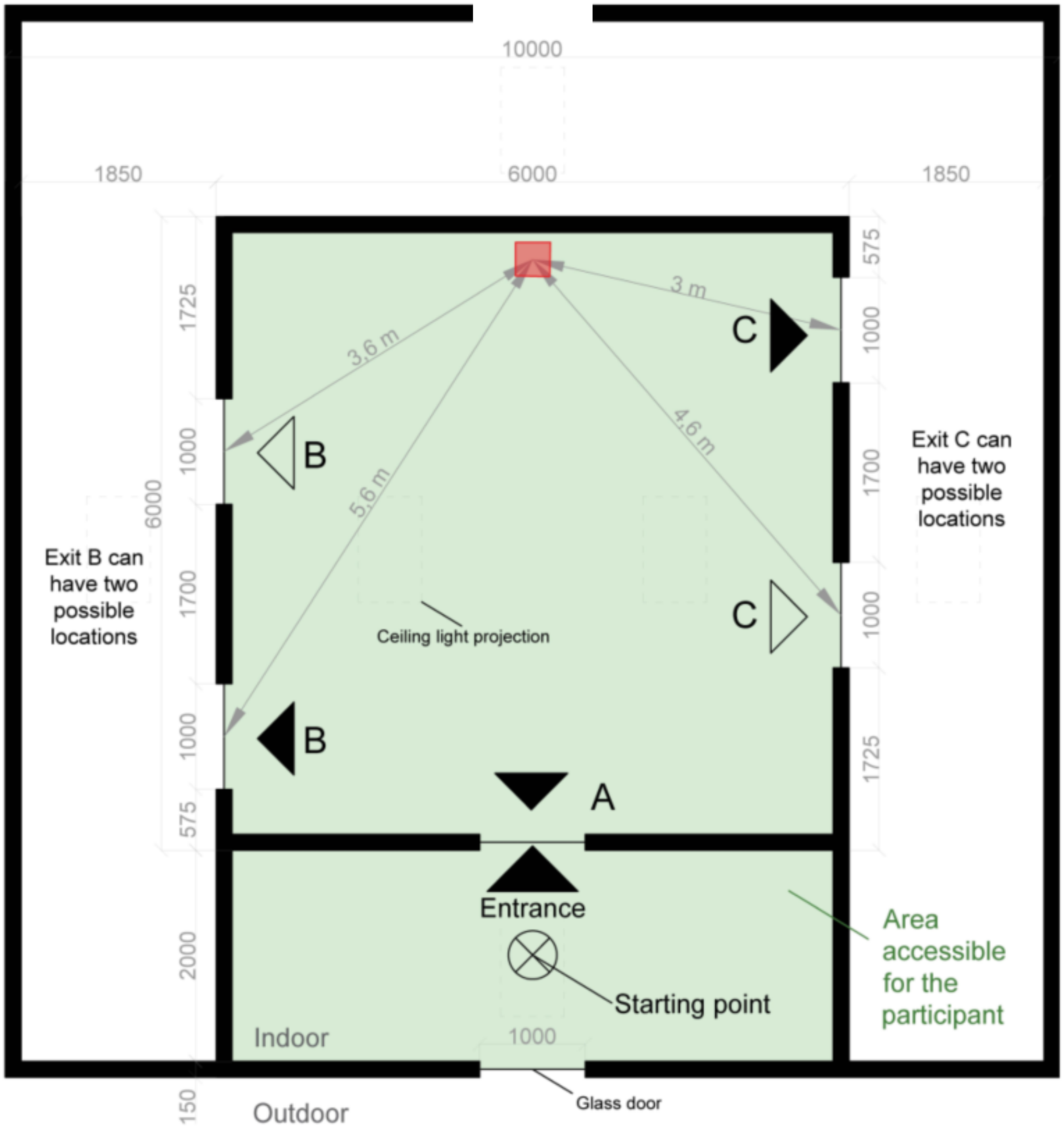

**Figure 1** – Geometry of the virtual environment.

Table 1 – Levels and values for each variable. Note: The values of the variable are in parenthesis

| Variable | Levels and Values | | |
|---|---|---|---|
| | **Exit A** | **Exit B** | **Exit C** |
| **NP** | 4<br>(0, 1, 5, 10) | 4<br>(0, 1, 5, 10) | 4<br>(0, 1, 5, 10) |
| **DIST** | 1<br>(6.0 m) | 2<br>(3.6 m, 5.6 m) | 2<br>(3.0 m, 4.6 m) |
| **SMOKE** | 2<br>(1, 0) | 2<br>(1, 0) | 2<br>(1, 0) |
| **FAM** | 1<br>(1) | 1<br>(0) | 1<br>(0) |

Due to the number of variables (4), the number of exits (3), and the possible dimensions of each variable, a high number of scenarios were available to investigate exit choice. The overall number of all the possible scenarios was 2048 (= $4^3$ x $2^2$ x $2^3$ x $1^3$). To identify the most useful scenarios to investigate exit choice, the *Efficient Design* approach in SP experimental design was used. The Efficient Design is a fractional factorial design used to identify an optimal subset of scenarios among all possible ones with the aim of maximising the amount of elicited information given a set sample size. In other words, compared to more traditional methods such as orthogonal design, efficient designs allow the analyst to obtain reliable estimates of the model coefficients with smaller sample sizes. This approach represents one of the best solutions when there is a large number of variables under investigation as well as when there is prior knowledge on how factors can affect the decision [47–49]. The Efficient Design used in this work is based on the minimisation of the *D-error* metric, which is the determinant of the asymptotic variance-covariance matrix to the power of 1/K, where K is the number of the parameters to estimate. In other words, the D-error metric is related to the p-values of the parameters to estimate with the data collected during the experiment. In fact, p-values are calculated using the variance matrix, which is the diagonal elements of variance–covariance matrix [47]. The Efficient Design approach generates an optimal solution when the researcher knows the expected values of the parameters which will be estimated with the experimental data. In this work, we used as priors the exit choice results published in [13].

In this work, we used Ngene [49] to run the Efficient Design, and to select the best eight scenarios to use for the exit choice experiments. The experimental scenarios are illustrated in Table 2.

Table 2 – Experimental scenarios identified using the Efficient Design

| Scenario | Exit A | | | Exit B | | | Exit C | | |
|---|---|---|---|---|---|---|---|---|---|
| | NP | DIST | SMOKE | NP | DIST | SMOKE | NP | DIST | SMOKE |
| 1 | 0 | 6 m | 0 | 10 | 3.6m | 1 | 5 | 4.6m | 1 |
| 2 | 5 | 6 m | 1 | 0 | 5.6 m | 1 | 10 | 3.0 m | 0 |
| 3 | 1 | 6 m | 1 | 1 | 5.6 m | 0 | 10 | 3.0 m | 0 |
| 4 | 10 | 6 m | 0 | 0 | 3.6 m | 0 | 1 | 4.6 m | 1 |
| 5 | 10 | 6 m | 0 | 1 | 3.6 m | 1 | 0 | 4.6 m | 0 |
| 6 | 5 | 6 m | 1 | 10 | 5.6 m | 0 | 0 | 3.0 m | 1 |
| 7 | 1 | 6 m | 1 | 5 | 5.6 m | 0 | 1 | 3.0 m | 0 |
| 8 | 0 | 6 m | 0 | 5 | 3.6 m | 1 | 5 | 4.6 m | 1 |

The scenario specifications in Table 2 were used to build the eight immersive VR experiences. These were developed using the Unity game engine [50]. The geometry of the digital building was created within Unity itself, and the furniture and doors were downloaded from the Unity asset store. The other evacuees in the room were simulated with Non-Playable Characters (NPCs). NPCs were developed in Adobe Fuse and animated using Mixamo [51]. The pipeline used to develop the NPCs is the one proposed in [52,53]. The smoke entering the room through the door during the fire emergency was generated using particle systems. Figure 2 shows some screenshots of the virtual experience. The VR experience was developed for a VIVE Pro headset by using the Steam VR packages available in the Unity asset store.

The application allowed the participants to visualise the digital scenario using the headset and to walk through it by walking in the physical space of 7m x 8m as shown in Figure 3. The movement of the participants in the physical space was tracked using four base stations located at the vertex of the rectangular tracking area (7 m x 8 m), as illustrated in Figure 3. This tracking area was large enough to allow participants to walk throughout the entire area highlighted in green in Figure 1.

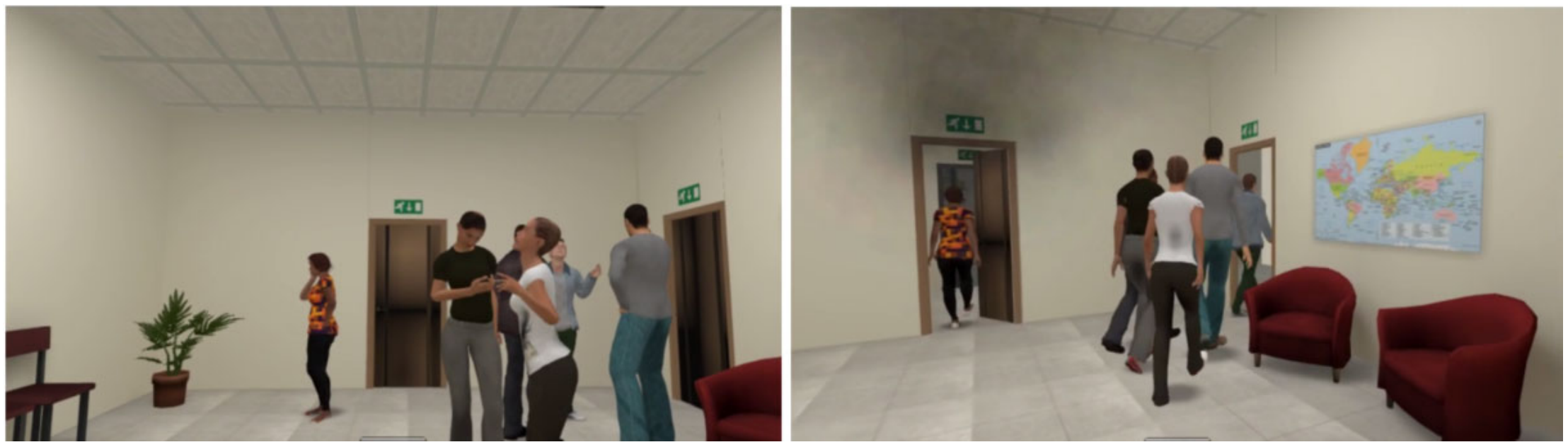

**Figure 2** – Screenshots of the virtual experience.

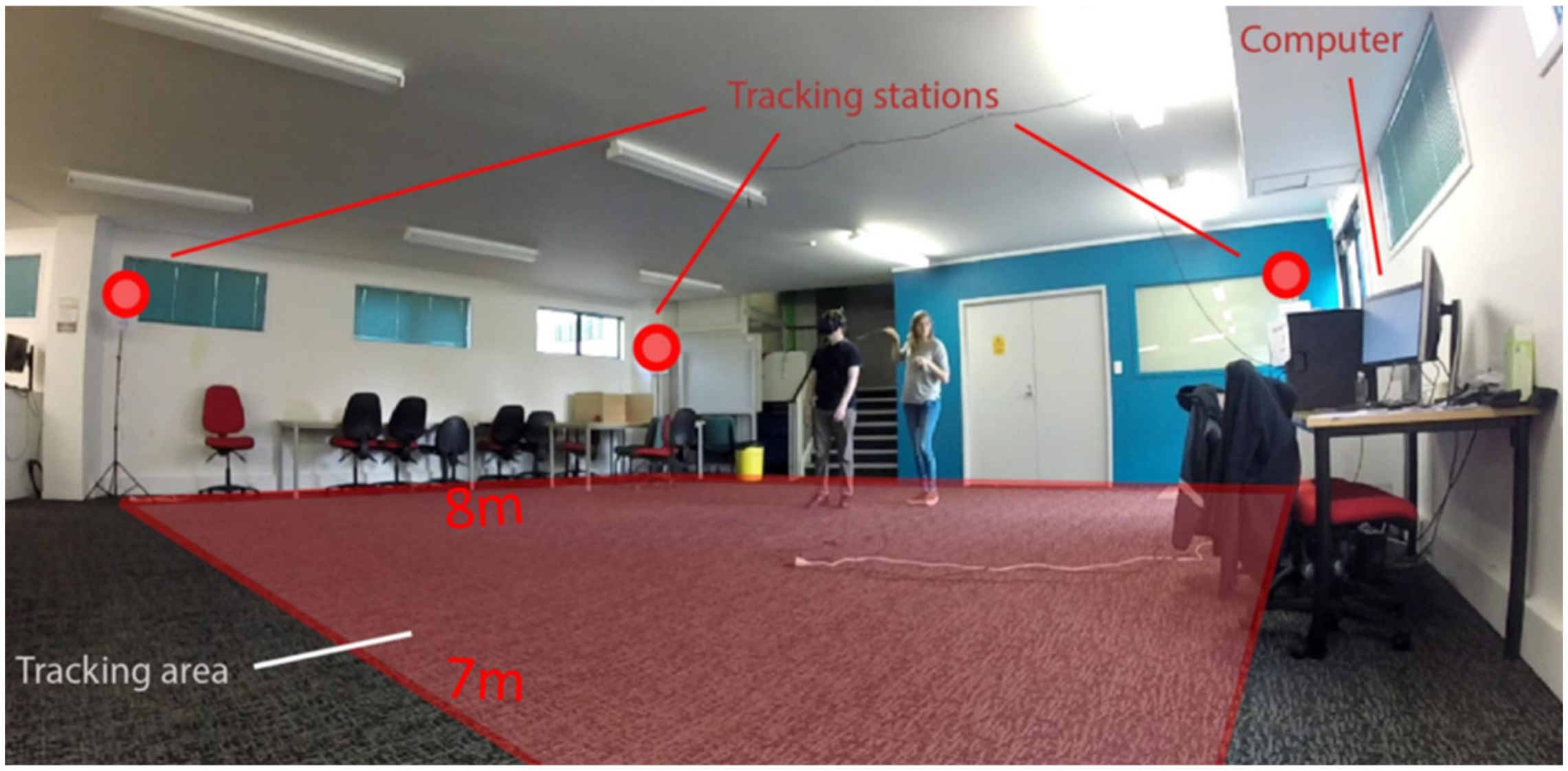

**Figure 3** – Physical space used to carry out the VR experiment

## 3.2 Experimental Procedure

The experiment took place in June 2019 at the Albany Campus of Massey University (New Zealand). The experiment included the following steps. All the participants were first asked to read and sign an approval form before taking part in the experiment, for ethical considerations and to make sure that none of them had any medical conditions that prevented them from taking part in the experiment. Then, a pre-experiment survey was administered to collect participants' demographics and their backgrounds regarding virtual reality and fire emergencies. After completing the questionnaire, the participants were asked to stand nearby the white door in Figure 3 and to wear the VR headset.

Each participant was asked to walk through an initial scenario that included only the geometry of the digital building used in the experiment. They were allowed to walk in the green areas shown in Figure 1 without walking through Exits B and C. This scenario gave participants the possibility to familiarise themselves with the environment and the VR navigation system before being immersed within the fire scenarios. As such, each participant used Exit A to enter in the virtual room. After walking in this room for 30-40s, the participant was asked to return to the starting position using Exit A (see Figure 1). This part of the experiment gave the participants the possibility to become familiar with Exit A.

Each participant was randomly assigned to four of the eight scenarios. The participants did not experience all the scenarios to avoid fatigue as the full experimental session aimed at not taking more of 30 minutes of their time. In each scenario, each participant was asked to enter the room using Exit A and to reach the red target shown in Figure 1 to attend a virtual meeting. The participants were not aware that the real intent of the experiment was observing their exit choice during an evacuation, as they were told that the VR experience was to demonstrate how VR can be used to run meetings. This allowed researchers to deceive the participants as they were not aware (at least for the first experiment) that a fire emergency would occur. A few seconds after the participants reached the red target, the fire alarm was activated, and the NPCs stopped previous activities immediately and started looking around before moving to a pre-computed exit following the numbers defined by Table 2. The pre-movement time varied for each NPC between 3 and 9 seconds. These reaction times were randomly assigned to the NPCs to avoid instances where NPCs could block each other from accessing the exits. When the alarm started, the smoke started entering the room through some of the doors depending on the scenario number (see Table 1). The experiment ended when the participants reached one of the three exits. Each exit choice made by the participants was recorded and stored in a local database (i.e., csv file). Similarly, their navigation path was monitored during their evacuation travel phase for each scenario.

After the VR experience, each participant was asked to fill a final questionnaire to collect feedback about the realism of the experiment, how easy it was to take part in the VR experience, their emotional states, the urgency perception and the validity of their behaviour. This was done by using seven-point Likert scale questions (-3= strongly disagree and +3= strongly agree). Each participant was asked to express their agreement/disagreement with the following statements:

- The virtual world was adequate/realistic (Realism 1);
- The virtual fire scenario was adequate/realistic (Realism 2);
- The interaction with other virtual people was adequate/realistic (Realism 3);
- I found running this VR scenario easy (Usability);
- This experience makes me feel scared/fearful (Emotion 1);
- Overall, this experience makes me feel tense/nervous (Emotion 2);
- Overall, this experience makes me feel anxious (Emotion 3);

- I felt the urgency to act/do something during the fire emergency (Urgency);
- I would act the same way in real life during a fire emergency (Validity).

Finally, each participant was asked to indicate which factors affected his/her choice with four open-ended questions (i.e., one question per choice). An open-ended response option was selected in this case to avoid influencing participants' answers with fixed options.

### 3.3 Participants

Most of participants were recruited through email, social networks or by other participants. Flyers were also distributed on the campus, and advertisements were published in the main buildings of Massey University. A total of 86 participants took part in the experiment. Most of them were staff or students of Massey University. The sample included 38 female and 48 male participants whose ages ranged from 18 to 66 years. The average age was 28.8 years; the standard deviation was 9.8 years while the $25^{th}$ and $75^{th}$ percentiles were 22 and 32 years, respectively. Participants' ethnicities varied; however, more than 47% were of Asian descent, and more than 30% defined themselves as European. The remaining 22 participants identified themselves as Middle Eastern (7%), African (5%) and from another ethnicity (10%).

### 3.4 Statistical Analysis

The choice data collected during the experiments were analysed using random utility models. In this work multiple forms of multinomial logit model were estimated. Random utility models rely on multiple assumptions [54,55]. The first assumption is that a decision-maker *q* assigns to each available choice alternative *i* a utility $U_{q,i}$. The utility is defined by a deterministic component $V_{q,i}$ and an error component $\varepsilon_{q,i}$ (Equation 1).

$$U_{q,i} = V_{q,i} + \varepsilon_{q,i} \quad \text{Equation 1}$$

In this work, we assume that the deterministic component has a linear specification (Equation 2).

$$V_{q,i} = \sum_{j} \beta_{i,j} X_{q,i,j} \quad \text{Equation 2}$$

where $X_{q,i,j}$ are the known values of the factors *j* perceived by the decision-maker *q* affecting the choice for the alternative *i*; and $\beta_{i,j}$ are parameters weighting the decision-makers' preferences related to the factors *j* (i.e., the parameters to estimate). It is possible to demonstrate that, when the error components are distributed as Extreme value Type I with variance $\pi^2/6$ and these distributions are independent and homoscedastic, the probability that the decision-maker *q* selects alternative *i* has the close formulation in Equation 3 (i.e., multinomial logit formulation).

$$P_{q,i} = \frac{\exp(V_{q,i})}{\sum_k \exp(V_{q,k})} \quad \text{Equation 3}$$

Equation 3 can be then used to build a likelihood function which is used to estimate the $\beta_{i,j}$ parameters by finding the combination of these parameters that maximises the likelihood function [56,57]. In this work, we estimated the multinomial logit models using the "mlogit" package available in R Studio [58].

Finally, the participants' responses to the final questionnaire assessing different aspects of the VR experience (see Section 3.2) were analysed by using boxplots to assess the average response and the variance of the answers. The open-ended responses on the factors affecting each of the four participants' choices were instead coded as follows: six binary variables were created representing the factors affecting choices:

- Follow NPCs;
- Avoid NPCs;
- Smoke;
- Distance;
- Familiarity;
- Other.

For each participant, a score of one was given for each of these factors if mentioned in their response.

## 4. Results

This section presents the results of the exit choice models proposed in this study in Section 4.1 while Section 4.2 provides the results of the participants' feedback regarding the virtual experience.

### 4.1 Exit Choice Models

In this work, we proposed two multinomial logit model formulations using the 344 choice observations collected from the experiment described in Section 3.

In *Model 1*, we estimated the parameters $\beta_{i,j}$ weighting the impact of NP, DIST, SMOKE and FAM using the specification in Equation 4. The parameters were all treated as generic across the three alternatives, meaning that we did not estimate alternative-specific parameters.

$$U_i = \beta_{np}\, NP_i + \beta_{dist}\, DIST_i + \beta_{sm}\, SMOKE_i + \beta_{fam}\, FAM_i$$

$$i = A, B, C \qquad \text{Equation 4}$$

The estimated parameters for Model 1 are shown in Table 3. The model shows that all the parameters were statistically different from zero, having their p-values below the level of significance of 0.05. The model also shows that participants tended to select the exit already selected by the NPCs exhibiting the impact of social influence or following behaviour. Further, the participants were more likely to choose the familiar Exit A while they were less likely to choose distant exits or exits having smoke.

Table 3 – Estimated parameters for Model 1

| Variable | Estimate | Std Error | z-value | P-value |
|---|---|---|---|---|
| $\beta np$ | 0.076 | 0.015 | 5.030 | 0.000 |
| $\beta dist$ | -0.378 | 0.079 | -4.771 | 0.000 |
| $\beta sm$ | -1.765 | 0.161 | -10.980 | 0.000 |
| $\beta fam$ | 0.795 | 0.206 | 3.864 | 0.000 |

*Model 2* was estimated to investigate the impact of participants' unawareness of the fire evacuation. In fact, as explained in Section 3.2, the participants were unaware that the first scenario would require them to evacuate, while they were aware of the evacuation when they repeated the experiment for the remaining 3 times. To assess if participants behaved differently when evacuating in the first scenario compared with the three that followed, we used a binary variable $C_1$, which equals 1 if the choice was made during the first evacuation or 0 otherwise. The specification of Model 2 is shown in Equation 5.

$$U_i = (\beta_{np} + C_1 \times \beta_{np1})\ NP_i + (\beta_{dist} + C_1 \times \beta_{dist1})\ DIST_i + (\beta_{sm} + C_1 \times \beta_{sm1})\ SMOKE_i + (\beta_{fam} + C_1 \times \beta_{fam1})\ FAM_i \quad \text{Equation 5}$$

$i = A, B, C$

The estimated parameters for Model 2 are shown in Table 4. The model shows that most of the parameters are statistically different from zero, having their p-values below the level of significance of 0.05. In line with Model 1, Model 2 shows that all the variables under investigation had an impact on the decision-making process. However, this second model shows that participants were more likely to choose exits used by other NPCs in their first choice as $\beta_{\text{np1}}$ *is* positive and significantly different from zero (i.e., $\beta_{\text{np}} + \beta_{\text{np1}} > \beta_{\text{np}}$). The model also shows that during the first evacuation the participants were still negatively affected by the smoke. In fact, they were less sensitive to it in the first evacuation as $\beta_{sm1}$ is positive and the sum of $\beta_{\text{sm}}$ and $\beta_{\text{sm1}}$ is negative (i.e., $\beta_{\text{sm}} + \beta_{\text{sm1}} < 0$).

Table 4 – Estimated parameters for Model 2

| Variable | Estimate | Std Error | z-value | p-value |
|---|---|---|---|---|
| $\beta np$ | 0.041 | 0.018 | 2.267 | 0.023 |
| $\beta dist$ | -0.439 | 0.094 | -4.688 | 0.000 |
| $\beta sm$ | -2.305 | 0.214 | -10.750 | 0.000 |
| $\beta fam$ | 0.735 | 0.248 | 2.968 | 0.003 |
| $\beta np1$ | 0.192 | 0.046 | 4.130 | 0.000 |
| $\beta dist1$ | 0.218 | 0.200 | 1.088 | 0.277 |
| $\beta sm1$ | 1.781 | 0.364 | 4.898 | 0.000 |
| $\beta fam1$ | 0.413 | 0.522 | 0.790 | 0.429 |

## 4.2 Sensitivity Analysis

A sensitivity analysis is conducted to show how the probability of choosing an exit can be affected by the variables investigated in Model 2 illustrated in Table 4. To simplify this analysis, we consider a scenario including only two exits (Exit A and Exit B).

To investigate the combined impact of the number of people using an exit (NP) and familiarity (FAM) in the exit selection, we assume that an evacuee is equally distant from Exit A and Exit B, there is no smoke located at either exit, and there are 5 evacuees using Exit B. Figure 4 shows the change of probabilities of the evacuees choice of Exit A for different numbers of people using this exit (i.e., NP for Exit A varies from 0 to 10 people). This investigation is run for three different scenarios: the evacuee is familiar only with Exit A (Fam A), only with Exit B (Fam B) and with both exits (Fam A&B). Figure 4 clearly shows that an increase in an increment of NP for Exit A leads to an increase in the probability of this exit of being selected. As an example, in the scenario in which the evacuee is familiar with both exits (Fam A&B), the probability of choosing Exit A increases from 0.23 to 0.76. Further, the familiarity of an exit generates a substantial shift in the probability curves. For instance, in the case of NP is equal to 5 for Exit A, the probability of choosing Exit A can range from 0.32 to 0.68 depending on the familiarity conditions.

The combined effect of distance (DIST) and familiarity (FAM) is also investigated in this section by assuming that there is no smoke located at either exit and there are equal numbers of people using both exits. The probabilities in Figure 5 are estimated assuming that an evacuee is 3 m distant from Exit B while the distance of Exit A varies from 0 m to 6 m. Once again, the analysis is run for three different scenarios: the evacuee is familiar only with Exit A (Fam A), only with Exit B (Fam B) and with both exits (Fam A&B). Figure 5 illustrates how distance has a strong impact on the probability of Exit A to be selected. In fact, considering the scenario Fam A&B, it is possible to observe that the probability of choosing Exit A decreases from 0.79 to 0.21 as distance increases. In line with the first sensitivity analysis, the familiarity of an exit generates a substantial shift in the probability curves in this case.

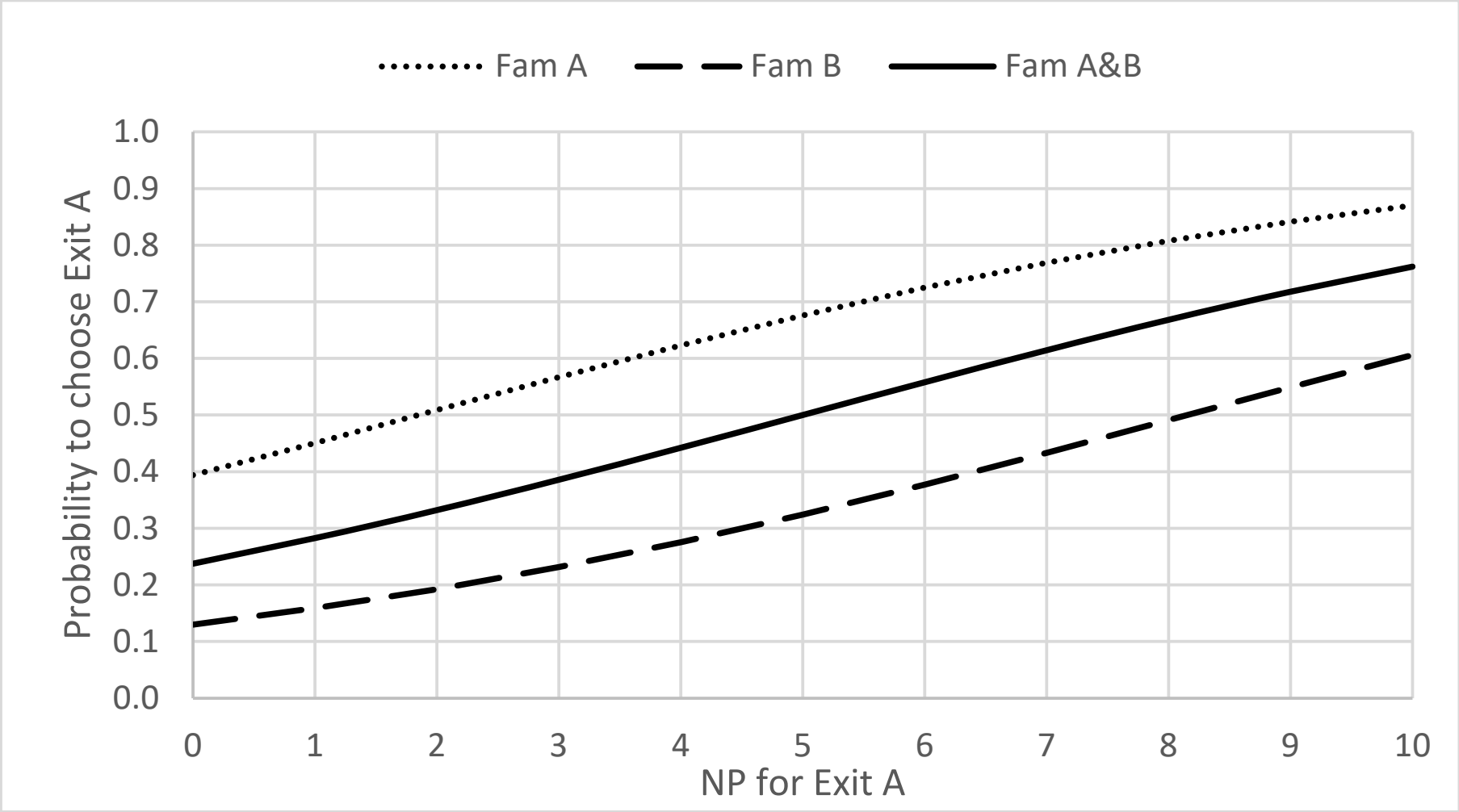

Figure 4 – Combined effect of NP and FAM on the exit choice

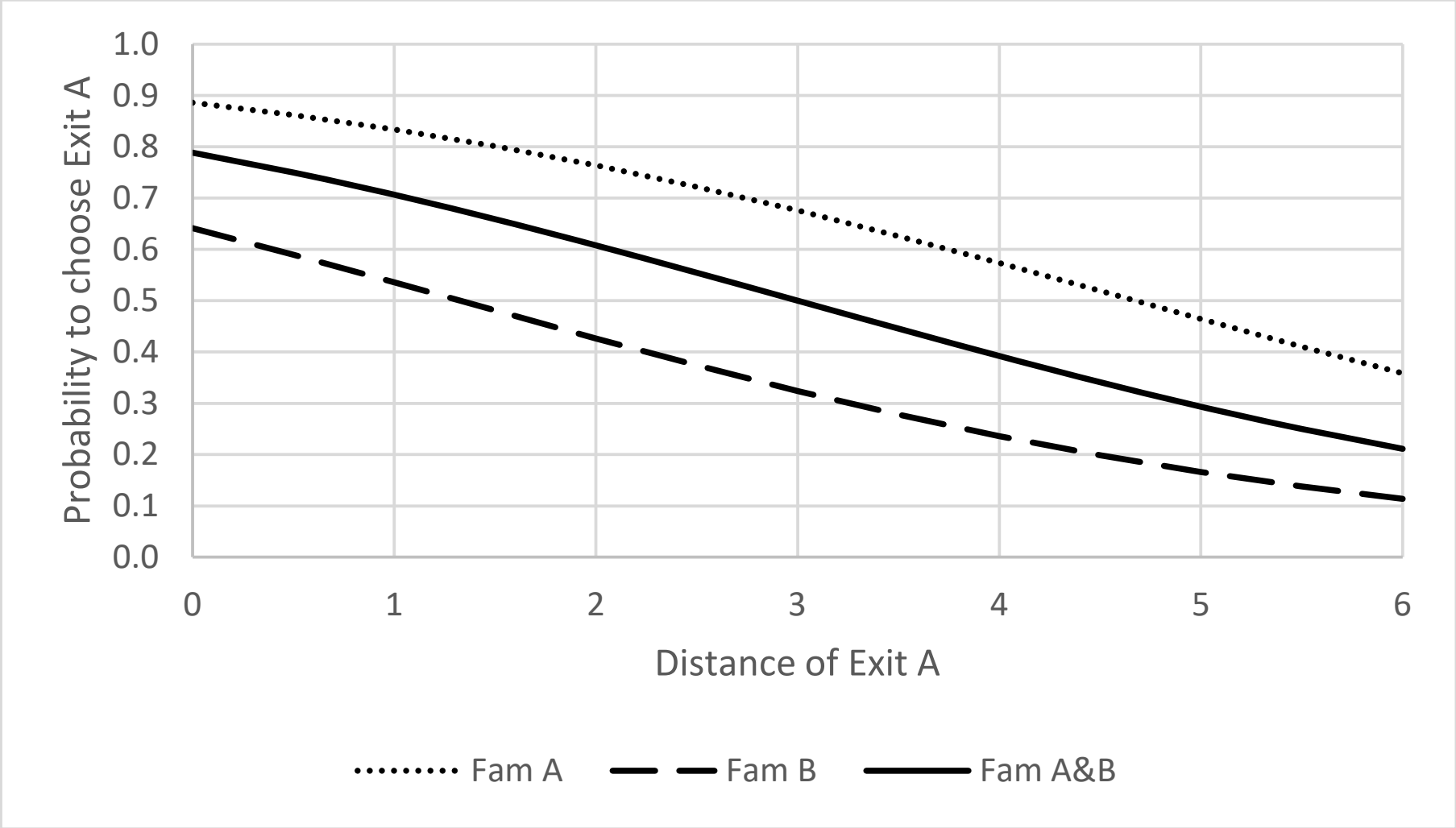

Figure 5 – Combined effect of DIST and FAM on the exit choice

It is worth highlighting that Figures 4 and 5 provide examples of possible analyses that can be carried out using the model in Table 4. The aim in this section is to demonstrate simple implementations of the proposed model that can be used to illustrate how particular factors affect the probabilities of evacuees choosing a specific door.

### 4.3 Respondents' Feedback

After the VR experience, participants were asked to provide feedback on the realism of the experiment, the engagement levels of the experiment, the ease with which they were able to take part to the VR experiment, and their emotional states (see Section 3.2). The feedback provided by the respondents is shown in the boxplots in Figure 6. The realism and the usability scores of the VR application are shown in Figure 6.a. The realism items demonstrate that participants found both the virtual world and the virtual fire realistic (see Realism 1 and 2). However, they stated that the NPCs were not as effective in terms of realism ratings. Further, Figure 6.a highlights that participants found the VR application easy to use.

Figure 6.b illustrates the scores for the emotional states of the participants, with Emotion 1 representing feelings of fear, Emotion 2 nervousness, and Emotion 3 anxiety. The majority of the sample did not report feeling any negative emotions. However, most of the participants reported that they felt urgency associated with taking action during the fire emergency and that they would behave in the same way in real life during the fire emergency.

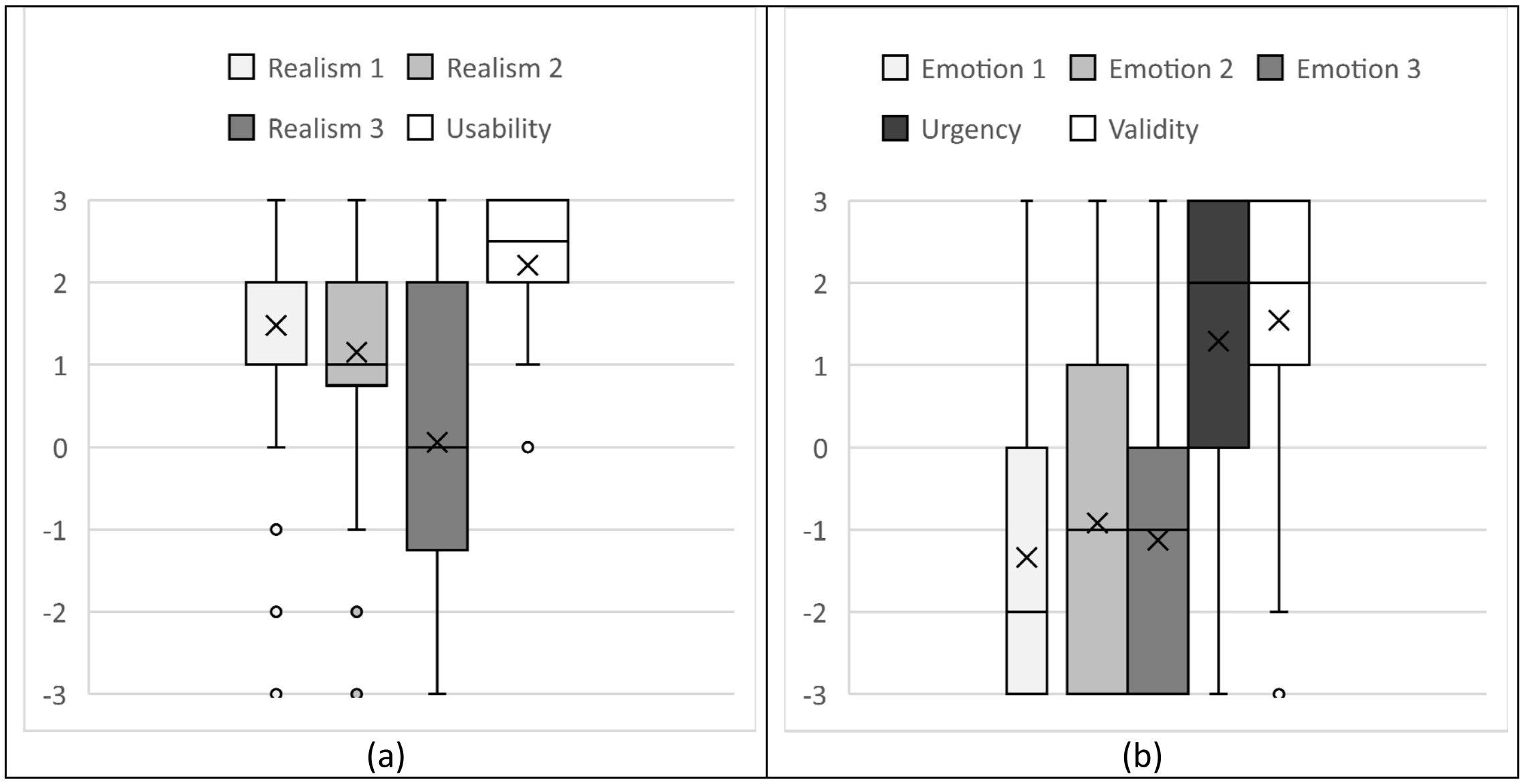

Figure 6 – Participants' scores on (a) realism and usability, and (b) their emotional state and perceived validity.

### 4.4 Reported Influential Factors of Exit Choice

After the VR experience, participants were asked to identify the factors affecting each of their four choices. The number of factors affecting the four choices is presented in Table 5. From this table, it is possible to observe that the large majority of the sample reported being affected by a single factor for all the four choices and that 20%-26% of the sample used a combination of two factors to select the exit. Only a few participants made their choices using three factors.

Table 5 – Number of factors affecting the choices

| | One Factor | Two Factors | Three Factors |
|---|---|---|---|
| **Choice 1** | 66<br>77% | 19<br>22% | 1<br>1% |
| **Choice 2** | 67<br>78% | 17<br>20% | 2<br>2% |
| **Choice 3** | 65<br>76% | 20<br>23% | 1<br>1% |
| **Choice 4** | 64<br>74% | 22<br>26% | 1<br>1% |

Table 6 illustrates which factors affected participants' choices. This table indicates that the most popular factor affecting the first choice was following NPCs (42%) while the second and third most popular factors were avoiding smoke (38%) and selecting the familiar exit (20%). Table 6 shows how these percentages change for the remaining choices 2-4; i.e., when participants were presumably more familiar with the experiment. In these cases, the most popular choice was smoke avoidance (49%-67%), while the number of participants who relied on following NPCs (16%-13%) decreased. This is in line with the modelling results reported in Table 4, showing that participants' decisions in the first experiment (first choice) were more sensitive to the behaviour of NPCs ($\beta np1>0$ and p-value <0.05) and less sensitive to the presence of smoke ($\beta sm1>0$ and p-value <0.05).

Table 6 – Factors affecting the choices

| | Follow NPCs | Avoid NPCs | Smoke | Distance | Familiarity | Other |
|---|---|---|---|---|---|---|
| **Choice 1** | 36<br>42% | 3<br>3% | 33<br>38% | 11<br>13% | 17<br>20% | 7<br>8% |
| **Choice 2** | 14<br>16% | 8<br>9% | 42<br>49% | 24<br>28% | 9<br>10% | 9<br>10% |
| **Choice 3** | 10<br>12% | 10<br>12% | 56<br>65% | 16<br>19% | 5<br>6% | 10<br>12% |
| **Choice 4** | 11<br>13% | 9<br>10% | 58<br>67% | 20<br>23% | 6<br>7% | 6<br>7% |

## Discussion and Conclusions

Models were developed that allowed for the prediction of exit choice based on four factors: the familiarity of the person with their exit of choice, the number of people using various exits, the presence of smoke at particular exits and the exit distance. Overall, the models' results showed that participants were influenced by all of the factors under investigation. Each of these results individually aligns with previous research in building fire evacuation.

First, our results involving familiarity provide support for Sime's affiliative model [29] and findings from studies of actual fire events [31], mock evacuation drills [33], and VR studies of building evacuations [18,34]. Also in line with Kinateder et al. [18] are our results (Figure 4) that show that the probability of someone using a familiar exit is enhanced with an increased number of other evacuees also using that exit. It should be noted that familiarity in this experiment was represented by a participant given the opportunity to virtually walk through that exit prior to the fire evacuation beginning. What that suggests is those building occupants may only need to have engaged with an exit once or twice to then feel greater confidence in using it to reach safety during a fire.

Our results also support much of the research on social influence with smaller crowds (~20 people or less), but not all. The findings from this study are similar to those from experimental building and tunnel studies with 1-3 participants, where people were more likely to follow another person to an exit (even if that exit was not identified as such) [38–40]. Additionally, similar to Kinateder and Warren [37] where participants were likely to follow crowds of 10 or less, this study found that people were likely to follow crowds as large as 20 people. However, our study contrasts with Lovreglio et al. [11,13] who, via online surveys using videos, found that participants mostly preferred less crowded exits. A similar contrast is observed in Haghani & Sarvi's [35] study conducted in field-type (non-virtual) settings that indicated participants tend to avoid movement in the same direction chosen by the majority. A possible reason for differences across these studies is the type of technology used to simulate the fire environment. While studies have shown that VR experiments like this one can provide similar results to those found in real-world settings [27,34,59], little work has been performed on the ecological validity of video stimuli (i.e. non-immersive experiments). As such, future studies are necessary to compare non-immersive and immersive VR settings.

This study also highlighted the impact of smoke on exit choice. Similar to [46], we found that the presence of smoke reduced the likelihood of choosing an exit. While previous research has shown that in some cases, people will walk through smoke, in this case, this study demonstrated that when people had other, seemingly safer options, they took them. This finding also suggests that smoky conditions reduce an exit's functional affordance, i.e., making it less likely to be perceived as an effective way to reach safety [42] and in turn, less likely to be used for evacuation.

This work also assesses the impact of asking the same participants to make choices in different scenarios. The technique of exposing the same participants to multiple-choice scenarios is a standard practice in transportation studies and was also adopted for exit choice studies (e.g., [11,13,49]); however the behavioral impact was not investigated previously. Our Model 2 in Table 4 is an attempt to assess if participants change their choice strategy by comparing the first choice they made with the remaining choices. The model shows that there was a significant change in the way other people and smoke affected the decisions. This was also confirmed by the qualitative analysis discussed in Section 4.4. The results in Table 6 highlight the impact of replication in experimental studies. These findings show that in their first scenario, when they were unfamiliar with the experiment, participants' exit choices were mainly affected by others in their environment. However, by the 2nd through 4th scenarios, the presence of smoke was more often cited as their factor of influence. In line with Model 2, these results suggest that at first, when participants were beginning to become familiar with the experiment, they relied more on others to help with their decision-making. As they became more familiar with the experiment and its scenarios, it is possible that participants focused more on the smoke as a hazard from which they needed to retreat. This “learning effect” has also been observed and investigated in other contexts of choice experiments (i.e., in more conventional non-VR formats). For example, Brouver et al. [60] reported in their empirical investigation that in a repeated choice experiment, learning did occur but that it did not exert any significant impact on econometric estimates. It would, however, be fair to assume that this observation may be case specific and not necessarily generalisable to all contexts and forms of choice experiments. It only shows that the effect of learning cannot be ruled out. In our case, being the participant unaware of an imminent evacuation for the first scenario (e.g., see the deception strategy in Section 3.2), it is possible to argue that the first choices made by the participants have higher ecological validity.

Results from this study also highlight the value that VR can bring to data collection of human behavior in building fires. Participant feedback rated the simulation of the “virtual world” and the fire environment as realistic. In addition to usability being rated as high, so were perceptions of urgency and the likelihood that they would act similarly in an actual fire emergency. Although these are participant perceptions and we lack actual data with which to compare, this feedback highlights the potential for continued use of VR in future behavioural studies. And as technology becomes more sophisticated and the virtual settings can increase in realism, researchers will be required to weigh the value of critical data with ethical practices when exposing participants to a scenario that might upset them, as discussed in [61].

The data presented in Table 5 shed light on an important phenomenon not necessarily captured by our logit models: an assessment of how many of the four factors played a role on each round of exit choice. In each scenario, the participants’ environment contained a combination of social influence, familiarity, exit distance and smoke factors, and one, a few, or all of these factors could have influenced their decision. Table 5 shows that a majority of participants identified only one factor as influential on their decision, with a smaller percentage listing two, suggesting that even with multiple factors present in the same environment, individuals may focus only on one to make their decision. However, the leading factors can be different among participants, as illustrated in Table 6, showing the heterogeneity of choice strategies. Decision scientists have argued that stressors, such as time pressures or uncertainty, can narrow a person’s perceptive field, causing them to pay attention to a select number of cues from their environment [62]. While this theory may hold here, additional research is required to further explore the dynamics at play.

This study has some limitations. The sample size of this studies is smaller than those used in previous studies. In addition, the participants of this study were all living in New Zealand. Previous online studies involved several hundreds of people or thousands of participants (e.g. [11,13]). However, the use of the Efficient Design mitigates this limitation as it allows researchers to obtain reliable estimates of the model coefficients with smaller sample sizes. Further, because of the sample size limit, this work does not investigate heterogeneity by using random parameter models or more advanced Bayesian hierarchical models as proposed recently by Song and Lovreglio [15]. As such, future studies are necessary to provide a bigger sample that can allow the investigation of exit choice heterogeneities.

The final limitation of this study is that several participants did not find the NPCs realistic enough. This might bias the way participants perceived NPCs and the social interaction observed in this study as predicted by the conceptual Threshold Model of Social Influence [63]. On the other hand, the self-reported ecological validity (see Figure 6.a) indicates that participants likely behaved in the experiment as they would in a real emergency. NPCs realism can be enhanced in future studies using the experimental deception proposed by Shipman et al. [64], who increased the believability of the NPCs in their terrorism experiments by making the participants believe that the NPCs, they view in the VR scenarios, were other participants taking part in the same experiment